\documentstyle[12pt]{article}

\title {Errata in \em {Enclyclopedia of Cosmology}}

\author {Kevin Krisciunas}
\date { }

\begin{document}
\maketitle

\vspace {-1 cm}

\begin {center}
     Joint Astronomy Centre \\
     660 N. A'ohoku Place \\
     University Park \\
     Hilo, Hawaii 96720 USA
\end {center}

\begin {abstract}
     I have noted a number of errors, most of them quite minor, in
the {\em Encyclopedia of Cosmology} (New York and London: Garland), 1993,
Norriss S. Hetherington, ed.  The majority occur in my mathematically
verbose article, ``Fundamental cosmological parameters".  Some errata were
passed on to me by Prof. Ralph Alpher.  In the interests of accuracy,
I feel that the corrections should be publicized, since no book review will
mention more than a couple of them.  Also, the incorrectly typeset equations
could lead to serious confusion in the minds of readers seeing such
material for the first time.
\end {abstract}

\newpage

\begin {center}
{\bf More serious errors}
\end {center}

\parindent=0pt

p. 223, right hand (RH) column, second to last equation (the definition
of $q_o$), the R-dot in the denominator should be squared

\vspace{5 mm}

p. 228, Eq. 45 should read the same as Eq. C9 on p. 244.  The expression
in square brackets should read

\begin{displaymath}
\left[ (1 + \Omega z)^{1/2} - 1 \right ]
\end{displaymath}

\vspace{5 mm}

p. 233, left hand (LH) column, line 14, for
``$29 \times 10^{-31}$" read ``$2-9 \times 10^{-31}$"

\vspace{5 mm}

p. 238, Eq. A4, last parenthetic expression, for
$\left( \frac {\Omega - 1} {\Omega} \right)$ read
$\left( \frac {\Omega - 2} {\Omega} \right)$

\vspace{5 mm}

p. 240, RH col, middle equation, for

\begin{displaymath}
\rm{sin}^{2/3}
\end{displaymath}

read

\begin{displaymath}
\rm{sinh}^{2/3}
\end{displaymath}

\vspace{5 mm}

p. 474, in Alpher and Herman's article on ``Origins of Primordial
Nucleosynthesis", equation 30 should be essentially the same as
Eq. A6 on p. 239.  Note that it should be ``arc cos", not ``arc cosh":

\begin{displaymath}
   (\Omega > 1):
 H_o T_o =  \frac {-1}{(\Omega - 1)} +
   \frac {\Omega}{2(\Omega -1)^{3/2}} \rm{arc cos} \left(
    \frac {2}{\Omega} - 1 \right)
\end{displaymath}

\vspace{5 mm}
\begin {center}
{\bf Less serious errors}
\end {center}

\vspace{5 mm}

p. 156, RH col, halfway down, it should read ``P. James E. Peebles"
without the comma after James

\vspace{5 mm}

p. 180, RH col, lines 1 and 2, for $\delta\lambda$ read $\Delta\lambda$

\vspace{5 mm}

p. 218, RH col, for ``$\Lambda = 21$ cm" read
``$\lambda = 21$ cm"

\vspace{5 mm}

p. 219, LH col, for ``If $\Delta \Lambda > 0$" read
``If $\Delta \lambda > 0$"; also, for ``$\Lambda = 6562.8$" read
``$\lambda = 6562.8$".

\vspace{5 mm}

p. 219, LH col, 3/4 down, it should read ``$\approx 1 + \beta +
{\cal O} (\beta^2)$", the last term meaning that it is ``of order"
$\beta^2$

\vspace{5 mm}

p. 221, LH col, halfway down, for ``t e L" read ``Let"

\vspace{5 mm}

p. 222, Eq. 15, eliminate comma

\vspace{5 mm}

p. 223, RH col, lines 1-2, for ``figure 3" read ``figure 4"

\vspace{5 mm}

p. 224, RH col, 3/4 down, for ``A = 13.0 Gyr" read ``$t_o$ = 13.0 Gyr"

\vspace{5 mm}

p. 223, RH col, middle equation, left hand side, for ``$\rho c$"
read ``$\rho_c$"

\vspace{5 mm}

p. 225, RH col, line 4, for ``Figure 4" read ``Figure 3"

\vspace{5 mm}

p. 226, LH col, 2/3 down, equation should read

\begin{displaymath}
 R^{1/2} dR = H_o dt
\end{displaymath}

\vspace{5 mm}

p. 228, RH col, line 1, for ``D $\rightarrow$ 0" read ``$\Omega
\rightarrow 0$"

\vspace{5 mm}

p. 229, LH col, second equation, all three terms in regular
parentheses should read ``(1 + z)".  The third one inadvertantly
has a minus sign in it.

\vspace{5 mm}

p. 230, RH col, last line, for

\begin{displaymath}
1 \propto d^{-2}
\end{displaymath}

read

\begin{displaymath}
l \propto d^{-2}
\end{displaymath}

\vspace{5 mm}

p. 231, Eq. 59, LH side, for ``$\Delta \rm{M_{bol}}$" read
``$\Delta \rm{m_{bol}}$"

\vspace{5 mm}

p. 234, RH col, 9 lines from bottom, for ``0.2/(or 0.1)"
read ``0.2 (or 0.1)"

\vspace{5 mm}

p. 239, RH col, second equation down, the typesetting is missing a long
vertical line, with the lower limit of the integration being 0

\vspace{5 mm}

p. 240, just before Eq. A11, for ``Hubble time" read ``Hubble times"

\vspace{5 mm}

p. 240, Eq. A11, LH side, for $\frac {t_o} {t_H}$ read
$\frac {t_o} {T_H}$

\vspace{5 mm}

p. 242, LH col, second to last equation, LH side is OK, but
on RH side there are no r-dots;
namely, the denominator should be
$(r^{\prime})^2$ and it should read $dr^{\prime}$ without the dot.

\vspace{5 mm}

p. 244, LH col, first line, for $2a/(1 + a)^2$ read $2a/(1 + a^2)$

\vspace{5 mm}

p. 302, RH col, at end, for ``light=years" read ``light-years"

\vspace{5 mm}

p. 471, LH col, 4 lines from bottom     delete comma after {\em k}

\vspace{5 mm}

p. 471, RH col, Eq. 7, for $\frac {\rho_r} {\rho_m}$ read
$\frac {p_r} {p_m}$

\vspace{5 mm}

p. 471, RH col, 4th line after Eq. 9, for ``2.735" read ``2.730"

\vspace{5 mm}

p. 471, LH col, immediately after Eq. 13b, for ``Where" read ``where"

\vspace{5 mm}

p. 473, RH col, last line, for ``if" read ``in"

\vspace{5 mm}

p. 643, RH col, the year of Tully and Fisher's article is 1977, not 1974

\end{document}